\documentclass[final,english]{bullsrsl}

\usepackage[latin1]{inputenc}

\usepackage[T1]{fontenc}

\usepackage{amssymb}
\usepackage{longtable}
\usepackage{subcaption}
\usepackage{graphicx}

\usepackage[numbers]{natbib}

\begin{document}
\title{A grid of Non-LTE line-blanketed atmosphere structures and synthetic spectra for subdwarfs}

%
%
%
%

\author[affil={1}, cofirst, corresponding]{Thayse A.}{Pacheco}
\author[affil={2}, cofirst]{Ronaldo S.}{Levenhagen}
\author[affil={1}]{Marcos P.}{Diaz}
\author[affil={1}]{Paula R. T.}{Coelho}

% ...

%
\affiliation[1]{Universidade de S\~ao Paulo, Instituto de Astronomia, Geof\'isica e Ci\^encias Atmosf\'ericas}
\affiliation[2]{Universidade Federal de S\~ao Paulo, Departamento de F\'isica}

\correspondance{thayse.pacheco@usp.br}

\date{31st October 2022}
\maketitle

\begin{abstract}
We present an update of the grid of detailed atmosphere models and homogeneous synthetic spectra for hot, high-gravity subdwarf stars. High-resolution spectra and synthetic photometry were calculated in the wavelength range 1,000 \texttt{\AA} -- 10,000 \texttt{\AA} using Non-LTE extensively line-blanketed atmosphere structures. 

\end{abstract}


\msccodes{65F15 65G50 15-04 15B99}

\begin{altabstract}
\textbf{Traduction du titre anglais en fran\c{}cais.} Nous pr\`esentons une mise \`a 
jour de la grille des mod\`eles d'atmosph\`ere d\'etaill\'es et des spectres synth\'etiques 
homog\'enes pour les \'etoiles sous-naines chaudes de haute gravit\'e. Les spectres haute 
r\'esolution et la photom\'etrie synth\'etique ont \'et\'e calcul\'es dans la gamme de longueurs 
d'onde 1.000 \texttt{\AA} - 10.000 \texttt{\AA} \`a l'aide de structures atmosph\'eriques 
non-LTE enti\'erement recouvertes de lignes.
\end{altabstract}

\altkeywords{spectres synth\'etiques, atmosph\`eres stellaires, 
profils de ligne, sous-naines}




\section{Introduction}

The formation mechanism and evolution of subdwarfs (sd) is yet to be fully understood. Hot sd  stars are thought to have evolved from low to intermediate-mass stars and reach far beyond the main sequence, at the blue end of the horizontal branch (HB) \citep{Heber2009}. While most of sds are found between the main sequence and the white dwarf (WD) regions in the HR diagram, some do overlap with WDs.
Perhaps this is due to the most accepted formation scenario being that these are extreme HB stars that have had their envelopes stripped through binary system interaction and are left with an exposed core \cite[e.g.][]{Geier2010}. This is supported by observations \cite{pelisoli2020}.

Homogeneous and widely available models for such objects are necessary in order understand the fundamental parameters and evolution of these unusual objects. As such, these must be computed using radiative transfer, non-local thermodynamic equilibrium (NLTE) physics, with extensively blanketed atmosphere structures \citep{metalNLTE1995, Lanz1997}. 

In \citep{pacheco} we present our original grid of model, covering eight temperatures within 10,000 $\le$ $T\textsubscript{eff}$ [K] $\le$ 65,000, three surface gravities in the range 4.5 $\le \log{g}$ [cgs] $\le$ 6.5, two helium abundances matching an extreme helium-rich and helium-poor scenarios for sds, and two metallicities representatives of the solar neighborhood ([Fe/H] = 0) and Galactic halo ([Fe/H] = -1.5 and [$\alpha$/Fe] = +0.4). In this work, we provide an update by adding 56 new models with three additional effective temperatures and two more surface gravity values, whilst still covering the same parameter space.

\section{Atmosphere Structure and Synthetic Spectra}

Atmospheric structure models in this work were computed using the \texttt{TLUSTY} code v205 and v208 \cite{hubeny1988, tlusty2011}, which calculates self-consistent solutions of radiative transfer and physical atmosphere states. Additionally, the most used stellar atmosphere models have been computed in LTE frameworks.  However, in the case of hot sds a NLTE computation has revealed significant differences that should be considered \cite[e.g.][]{metalNLTE1995}, and that approach was taken in this work. A full description of this methodology can be found in \citep{pacheco}. 

As an addition to that work, here we expand our grid by including sd-structure models composed of one additional surface gravity point ($\log{g}$ [cgs]: 5.0) for five effective temperatures ($T\textsubscript{eff}$: 15,000, 20,000, 25,000, 30,000 and 35,000 K). Moreover, we computed a new set with three effective temperatures ($T\textsubscript{eff}$: 17,500, 22,500 and 27,500 K), each one with three surface gravities ($\log{g}$ [cgs]: 4.5, 5.0, and 5.5) and four different chemical abundances: solar and He-rich, solar and He-poor, halo and He-rich, halo and He-poor. The numerical He-rich and He-poor variations were chosen based on linear fits as a function of temperature based on \cite{Lei2019} as shown in table 1 for the new models. See table 2 from section 2 of \citep{pacheco} for the He abundances as a function of temperature originally used and not show here.

\begin{table}[htb]
\centering
\caption[]{Numerical He Abundances \footnotemark
} 
\begin{tabular}{ccc}
\hline \hline
$T\textsubscript{eff}$ [K]  & log($n$\textsubscript{He}/$n$\textsubscript{H}) \textsubscript{poor}  & log($n$\textsubscript{He}/$n$\textsubscript{H}) \textsubscript{rich}       \\
\hline 
17,500 & -4.05  & -2.80  \\
22,500 & -3.41  & -2.12   \\
27,500 & -2.78  & -1.45  \\
 \hline 
\end{tabular}
\label{n_He}
\end{table}
\footnotetext{for the remaining He abundances please see table 2 of \cite{pacheco}.}

We computed the grid of spectra using the \texttt{SYNSPEC} code which is designed to synthesize the spectrum from atmosphere model structures. We used NLTE assumption based on \texttt{TLUSTY} models, using reference atomic line lists from \cite{Coelho2014} and with special treatment of the H and He lines computed using tables from \cite{Tremblay2009}. This method is identical to that described in \cite{pacheco}, where examples of sample spectra can be found and comparisons between existing models for sample targets are found to be in good agreement (see section 3 of \citep{pacheco} for more details). 

\begin{figure}[h!]
\centering
\includegraphics[width=.8\linewidth]{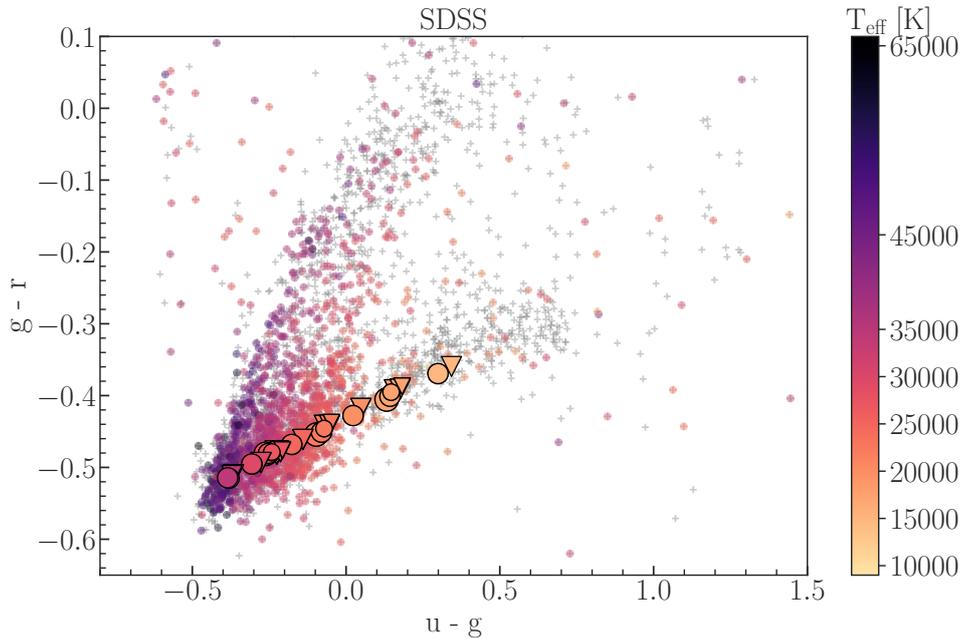}
\caption{Color-color diagram (\textit{g - r} vs. \textit{u - g}) composed of observational data on subdwarfs from SDSS (see text) with a hue scale and gray crosses representing determined and undetermined effective temperatures, respectively. The new synthetic colors (wiht black contours) from our grid are plotted with the same hue scale as observational data showing effective temperatures; 
circles indicate solar abundances and triangles the low halo metallicity. \label{sdss}}
\end{figure}

With these models in place, we are able to compute predicted magnitudes for several standard photometric filters and compare with observations. Figure \ref{sdss} shows a color-color diagram of the composed observational data on sds from SDSS from Geier catalog \citep{Geier2010} where we include our synthetic colors from \citep{pacheco}, along with updated results from the present work in the same hue, size and scale for comparison. 

\section{Conclusions}

In this manuscript we present an extension to our grid of theoretical spectra computed with NLTE, based on extensively line blanketed models for 
sd stars and compare it with photometric observations. We find good agreement between modeled and observed colors.

\begin{acknowledgments}
This study was financed in part by the Coordena\c{c}\~ao de Aperfei\c{c}oamento de Pessoal de N\'ivel Superior - Brasil (CAPES), Finance Code 001.
\end{acknowledgments}


\begin{furtherinformation}

\begin{orcids}
%
\orcid{0000-0002-8139-7278}{Thayse Adineia}{Pacheco}
\orcid{0000-0003-2499-9325}{Ronaldo Savarino}{Levenhagen}
\orcid{0000-0002-6040-0458}{Marcos Perez}{Diaz}
\orcid{0000-0003-1846-4826}{Paula R. T.}{Coelho}
\end{orcids}




\end{furtherinformation}


%

\bibliographystyle{bullsrsl-numen}


\begin{thebibliography}{10}
\providecommand{\url}[1]{#1}
\providecommand{\urlprefix}{URL }

\bibitem{Heber2009}
Heber, U. (2009) {Hot Subdwarf Stars}.
\newblock \araa, 47(1), 211--251.

\bibitem{Geier2010}
{Geier}, S., {Heber}, U., {Podsiadlowski}, P., {Edelmann}, H., {Napiwotzki},
  R., {Kupfer}, T. and {M{\"u}ller}, S. (2010) {Hot subdwarf stars in close-up
  view. I. Rotational properties of subdwarf B stars in close binary systems
  and nature of their unseen companions}.
\newblock \aap, 519, A25.

\bibitem{pelisoli2020}
{Pelisoli}, I., {Vos}, J., {Geier}, S., {Schaffenroth}, V. and {Baran}, A.~S.
  (2020) {Alone but not lonely: Observational evidence that binary interaction
  is always required to form hot subdwarf stars}.
\newblock \aap, 642, A180.

\bibitem{metalNLTE1995}
{Lanz}, T. and {Hubeny}, I. (1995) {Non-LTE Line-blanketed Model Atmospheres of
  Hot Stars. II. Hot, Metal-rich White Dwarfs}.
\newblock \apj, 439, 905.

\bibitem{Lanz1997}
Lanz, T., Hubeny, I. and Heap, S.~R. (1997) Non-{LTE} line-blanketed model
  atmospheres of hot stars. {III}. hot subdwarfs: The {sdO} star {BD}
  +75$^o$325*. 
\newblock \apjl, 485(2), 843--858.

\bibitem{pacheco}
{Pacheco}, T.~A., {Diaz}, M.~P., {Levenhagen}, R.~S. and {Coelho}, P. R.~T.
  (2021) {A Grid of Synthetic Spectra for Subdwarfs: Non-LTE Line-blanketed
  Atmosphere Models}.
\newblock \apjs, 256(2), 41.
\newblock \url{https://doi.org/10.3847/1538-4365/ac2508}.

\bibitem{hubeny1988}
Hubeny, I. (1988) A computer program for calculating non-lte model stellar
  atmospheres.
\newblock Computer Physics Communications, 52(1), 103 -- 132.

\bibitem{tlusty2011}
{Hubeny}, I. and {Lanz}, T. (2011).
\newblock {TLUSTY: Stellar Atmospheres, Accretion Disks, and Spectroscopic
  Diagnostics, Astrophysics Source Code}.

\bibitem{Lei2019}
Lei, Z., Zhao, J., N{\'{e}}meth, P. and Zhao, G. (2019) { New Hot Subdwarf
  Stars Identified in Gaia DR2 with LAMOST DR5 Spectra. II. }.
\newblock \apj, 881(2), 135.

\bibitem{Coelho2014}
Coelho, P. R.~T. (2014) {A new library of theoretical stellar spectra with
  scaled-solar and $\alpha$-enhanced mixtures}.
\newblock \mnras, 440(2), 1027--1043.

\bibitem{Tremblay2009}
Tremblay, P.~E. and Bergeron, P. (2009) Spectroscopic analysis of da white
  dwarfs: Stark broadening of hydrogen lines including nonideal effects.
\newblock \apj, 696, 308--322.

\end{thebibliography}

\end{document}